%% file: paper.tex
\documentclass[iop]{emulateapj}

\usepackage{natbib}
\usepackage{amsmath}
\usepackage{threeparttable}
\usepackage{amssymb}
\usepackage{graphicx}
\usepackage{longtable}
\usepackage{appendix}
\usepackage{float}
\usepackage{hyperref}

\renewcommand{\t}[1]{\mathrm{#1}}
\renewcommand{\H}{\ensuremath{\mathrm{H}}}

\newcommand{\msun}{\ensuremath{M_\odot}}

\newcommand{\sunits}{\mbox{\msun ~pc$^{-2}$}}

\newcommand{\x}[1]{\ensuremath{X_{\mathrm{#1}}}}

\newcommand{\e}{\ensuremath{\mathrm{e}}}

\newcommand{\co}[1]{\mbox{$^{#1}$CO}}

\newcommand{\av}{\ensuremath{\mbox{$A_{\rm V}$}}}

\renewcommand{\d}[1]{\ensuremath{\mbox{d}{#1}}}

\defcitealias{Solomon_1987}{S87}
\defcitealias{Bolatto_2008}{B08}

\shortauthors{Imara \& Forbes}

\begin{document}

\title{The Global Structure of Molecular Clouds: I. Trends with Mass and Star Formation Rate}

\author{Nia Imara}
\affiliation{University of California, Santa Cruz, 1156 High Street, Santa Cruz, CA 95064}
\email{nimara@ucsc.edu}

\author{John C. Forbes}
\affiliation{School of Physical and Chemical Sciences--Te Kura Mat\=u, University of Canterbury, Christchurch 8140, New Zealand}
\affiliation{Center for Computational Astrophysics, 162 5th Avenue, New York, NY 10010}

\begin{abstract}
We introduce a model for the large-scale, global 3D structure of molecular clouds. Motivated by the morphological appearance of clouds in surface density maps, we model clouds as cylinders, with the aim of backing out information about the volume density distribution of gas and its relationship to star formation. We test our model by applying it to surface density maps for a sample of nearby clouds and find solutions that fit each of the observed radial surface density profiles remarkably well. Our most salient findings are that clouds with higher central volume densities are more compact and also have lower total mass. These same lower-mass clouds tend to have shorter gas depletion times, regardless of whether we consider their total mass or dense mass. Our analyses lead us to conclude that cylindrical clouds can be characterized by a universal structure that sets the timescale on which they form stars.
\\
\end{abstract}

\section{Introduction}
Where, when, and how stars form is intimately tied to the physical structure of their birth environments, molecular clouds. Understanding the structure of clouds is key to developing a complete picture of star formation and, ultimately, galaxy evolution.  

It has been known for some time that molecular clouds have certain global properties in common.  In his seminal study, \citet{Larson_1981} measured the properties of nearby molecular clouds using CO observations and found that they follow three empirical scaling relationships: (1) the velocity dispersion of clouds increases with size, according to $\sigma\propto R^{0.5}$; (2) clouds are in an approximate state of virial equilibrium, with $5\sigma^2_v R/GM \sim 1$; and (3) molecular cloud mass and size follow a power-law scaling, $M \sim R^2$.  

The third relation suggests that molecular clouds have constant surface density.  Using dust extinction instead of CO observations to determine the clouds masses and sizes enabled \citet{Lombardi_2010} to investigate the validity of Larson’s third “law” over a larger dynamical range of column densities.  They found that molecular clouds have column density variations spanning more than two orders of magnitude, though the clouds' average column densities above a given extinction threshold are nearly constant.  From the theoretical perspective, \citet{Ballesteros-Paredes_2012} demonstrated that the roughly constant surface density of clouds is a consequence of observational methods that tend to define clouds at an extinction threshold close to the peak of the column density probability distribution function, which falls off rapidly at higher column densities.  

In this paper, we start from the premise that molecular clouds may \textit{not} necessarily have constant surface density. We explore the idea that variations in the surface density of clouds may contain information about their evolutionary state and their capacity to form stars. In other words, we ask the question: what is the global structure of molecular clouds, and how does this structure relate to star formation?

To address this question, we propose a simple three-dimensional model for cylindrical molecular clouds and apply it to dust extinction measurements of a sample of nearby clouds having a range of star formation rates (SFRs). The model is inspired by observations showing that on global scales, clouds frequently have elongated or filamentary shapes, i.e., they have high aspect ratios.

The molecular interstellar medium is pervaded by filamentary structures at a range of spatial scales, including the scale of molecular clouds ($\sim 10 - 100$ pc) \citep[e.g.,][]{Andre_2010, Molinari_2010, Li_2013, Schisano_2020, Wang_2020, Zucker_2018, Zhang_2019}. Using \co{13} data taken with APEX, \cite{Neralwar_2022} identified a population of more than 10,000 molecular clouds in the inner Galaxy and classified them according to morphology. They concluded that most of the clouds in their sample are elongated (57\%), with the remaining belonging either to the ``ring-like," ``concentrated," or ``irregular" classes.

In this work we consider elongated clouds.  (To avoid confusion regarding the filaments existing \textit{within} clouds as part of their substructure, we will use the terms ``elongated" or ``cylindrical" when referring to the global morphology of clouds.)  Our key results are that (1) this is a useful model for understanding the architecture of clouds in a way that takes into account their three-dimensional nature, and (2) clouds can be characterized by a universal structure that links to the efficiency with which they form stars.

\begin{figure}
\epsscale{.70}
\plotone{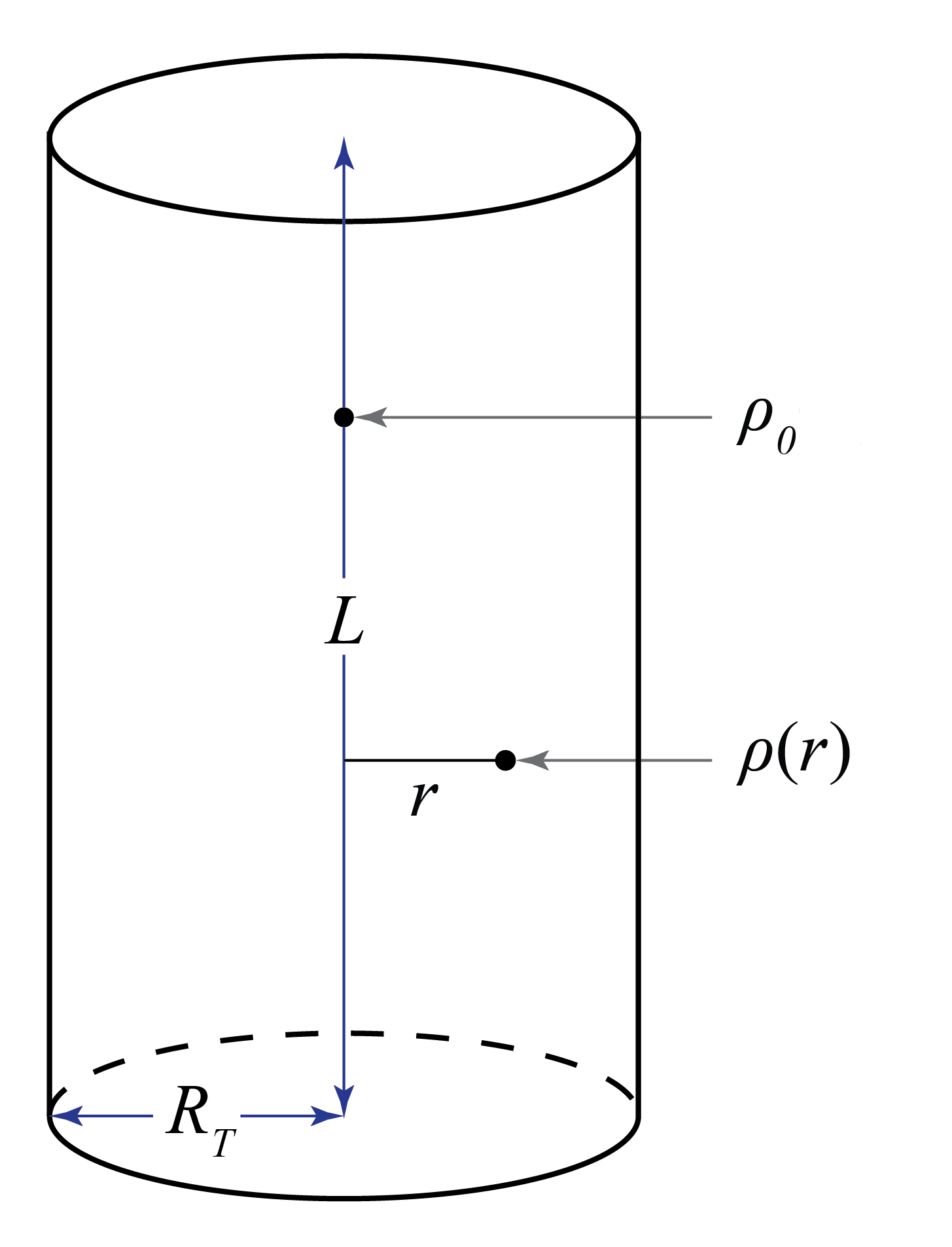}
\caption{Illustration of our simple cylinder model. Parameters include: the cloud length along the spine ($L$), the fixed maximum radius ($R_T$), the density along the spine ($\rho_0$), and the variable density $\rho(r)$ at some distance $r$ from the spine.  \label{fig1}}
\end{figure}

\section{The Model}\label{sec:model}

Surface density maps of Galactic molecular clouds show that they often have elongated shapes, with much of their high-column-density gas concentrated along a central axis. For such clouds, one might reasonably assume---as we do here---that this elongated appearance is because their 3D morphology is cylindrical. Some computer simulations show that structures appearing to be filamentary/cylindrical are sometimes more flattened, sheet-like structures viewed in projection \citep[e.g.,][]{Imara_2021}. Recent observational work by \cite{Tritsis_2022}  and \cite{Rezaei_2023} to infer the volume density structure of molecular clouds using dust maps support this view. In this study, we model molecular clouds as cylinders and choose a sample clouds that appear to approximate this morphology. We will consider other geometries in future work.

One of our goals is to tease out information about the volume density distribution of gas that has been projected onto the plane of the sky in surface density maps. Rather than trying to capture all of the intricate substructure known to comprise molecular clouds---e.g., filaments, clumps, and cores---our goal is to characterize their large-scale, global structure.

Figure \ref{fig1} shows a labelled illustration of our model. We call the length of a cloud $L$. In surface density maps of clouds like the Orion Molecular Clouds and the Perseus Molecular Cloud, you often see that much of the highest column density gas is concentrated along a central axis \citep[e.g.,][]{Lombardi_2010, Lombardi_2011, Imara_Burkhart_2016}---what we will refer to here as the \textit{spine}. The perpendicular distance from the spine to any point is $r$, and the maximum radius $R_T$ is the distance from the spine to the edge. 

We define the radial volume density profile of a cloud, $\rho(r)$, as 
\begin{equation}\label{eq:rho1}
\rho =\frac{\rho_0}{ \left[ 1 + \left( \frac{r}{r_0} \right)^2 \right]^\alpha  },
\end{equation}
where $\rho_0$ is the peak density along the central spine of the cloud, and  $r_0$ is the ``turnover'' radius inside of which the density is roughly constant, i.e.,  $\rho \approx \rho_0$. The exponent $\alpha$ describes how quickly the density falls off beyond $r_0$. Going forward, we refer to the volume density profile simply as $\rho$ and consider the radial dependence to be implied. 

The expression for an isothermal cylinder, in which $\alpha =2$, was derived by \citet{Stodolkiewicz_1963} and \citet{Ostriker_1964}: $\rho_{\rm iso}=\rho_0/[1 + r^2/8r_0^2]^2$.  In this instance, the scaling radius, $r_0$, is equivalent to the Jeans length: $r_0 = \sqrt{c_s^2/4\pi G\rho_0}$, where $c_s$ is the isothermal sound speed.  Clouds with the $\alpha =2$ profile would have a steeply declining density of $\rho\propto r^{-4}$ far from the central axis.  In observed cylindrical clouds, however, the density appears to falls off more moderately \citep[e.g.,][]{Fiege_2000, Arzoumanian_2011, Toci_2015}.  In our model, we leave $\alpha$, $\rho_0$, and $r_0$ as free parameters.

Our goal is to apply this model to real clouds and fit the free parameters, $\alpha$,  $\rho_0$, and $r_0$.  To do so, we must derive the two-dimensional mass surface density. The surface density at each projected distance $b_\perp$ from the cylinder's axis of symmetry, $\Sigma_{\rm model}(b_\perp)$, is given by integrating all volume densities along the line-of-sight:
\begin{equation}\label{eq:surf}
\Sigma_{\rm model}(b_\perp) = 2\int_{b_\perp}^{R_T} \rho(r) \frac{r \d{r}}{\sqrt{r^2 -b_\perp^2} },
\end{equation}
where $r$ still refers to the 3D distance from the symmetry axis. The result of the integral can be expressed with the Gauss hypergeometric function 2F1 to obviate the need for numerical integration. Note that any variation of $\rho_0(L)$, that is the density along the spine, is averaged out in the data (see Equation \ref{eq:colavg}), so there is no reason to include it in the model. Moreover for simplicity we assume that $r_0$ does not vary along the symmetry axis. These are deliberate choices to compare the model and data in a space where the model has a chance of being able to fit the data well. Similarly any inclination of the cloud, to a good approximation, would only change the overall density normalization. When we apply Equations \ref{eq:rho1} and \ref{eq:surf} to the data, we are therefore fitting for $\rho_0/\cos~i$ rather than $\rho_0$ itself, where $i$ is the angle between the plane of the sky and the cylinder's symmetry axis.

In the next section, we describe the observations and how we apply the model to them.

\section{Observations}\label{sec:observations}

\subsection{Cloud sample}
We built our sample of clouds (Table 1) based on their appearance in surface density maps and their proximity. All reside in the Solar Neighborhood at distances less than 1400 pc, which affords us good spatial resolution. To test our model, we used maps created from observations of dust extinction.

We generated the extinction maps using the near-infrared color excess method of \cite{Lombardi_2009, Lombardi_2011}. This technique uses JHK photometric data from the Two Micron All Sky Survey \citep{Skrutskie_2006} to measure dust extinction along many lines of sight towards clouds. For each individual cloud, measuring the extinction involves the selection of a control field where the extinction is negligible to calibrate the intrinsic colors of the stars. The resulting maps in units of visual extinction, ${A}_{{\rm{V}}}$, can be converted into units of surface density. Lombardi offers an interactive website called iNICEST that we used to construct the clouds, based on our chosen parameters.\footnote{http://interstellarclouds.fisica.unimi.it/html/index.html}  The resulting maps are shown in Figure \ref{fig:mosaic}.

\subsection{Cloud properties}
Cloud properties are extracted from the extinction maps as follows. First, a boundary in galactic latitude and longitude is chosen for each cloud (see Table 1), and a particular distance from the Sun to the cloud is adopted. Next a straight line in latitude and longitude is chosen to represent the cloud's spine or symmetry axis. This is the line that minimizes the L1 distance between the line and pixels with $A_V>A_{V,\mathrm{high}}$ in the map. $A_{V,\mathrm{high}}$ is the $\sim 99$th percentile of the distribution of $A_V$'s greater than 1. Once the spine is defined, we define a series of bins in the quantity $b_\perp$, the perpendicular distance away from the spine in the plane of the sky in pc. To do so\footnote{The distances do not matter for deriving an average column density within angular bins on the sky, but because we use bins of physical distance for $b_\perp$, which pixels are assigned to which bins does depend on distance. Distances also enter explicitly when deriving cloud masses.} we adopt distances from \citet{Zucker_2020} based on dust extinction of stars with distances measured from Gaia EDR3: distances for lines of sight from their catalogue that fall within the $A_V>1$ footprint of each cloud are averaged. 

Any pixels in the map whose centers fall within a particular bin are averaged, producing a sequence of column densities $\Sigma_i$ associated with bin centers $b_{\perp,i}$:
\begin{equation} \label{eq:colavg}
    \Sigma_i = \left(\sum_j \mathcal{C} A_{V,j} \mathcal{I}_{b_\perp, ij}  \right) \Big/ \left(\sum_j \mathcal{I}_{b_\perp, ij} \right)
\end{equation}
where $A_{V,j}$ is the $A_V$ value of pixel $j$, and $\mathcal{I}_{b_\perp,ij}$ is an indicator function,
\begin{equation}
    \mathcal{I}_{b_\perp, ij} = \begin{cases}
1    &   \text{if}~ b_{\perp,i} - \frac12\Delta b_\perp \le b_{\perp,j,\mathrm{pix}} < b_{\perp,i}+\frac12\Delta b_\perp, \\
0    &   \text{if}~\mathrm{otherwise}.
\end{cases}
\end{equation}
where $b_{\perp,j,\mathrm{pix}}$ is the $b_\perp$ coordinate of the $j$th pixel. The bin width $\Delta b_\perp$ is chosen to be 5 times the pixel scale of each map at the median adopted distance of each cloud to minimize errors from using the pixel centers to assign pixels to bins (as opposed to computing intersections of bin edges with individual pixels). The conversion factor from $A_V$ to a surface density in mass per unit area is \citep{Guver_2009}
\begin{equation}
\mathcal{C} = 2.21 \cdot 10^{21} \frac{\mathrm{H}\ \mathrm{nuclei}\ \mathrm{cm}^{-2}}{A_V\ (\mathrm{mag})} \cdot m_\mathrm{H},
\end{equation}
where $m_\mathrm{H}$ is the mass of a Hydrogen atom. Note that the maps themselves are in units of $K$-band magnitudes of extinction, $A_K$. We take $A_K/A_V = 0.11$.

The $\Sigma_i$ profiles extend out to $b_{\perp,\max}$ the value of $b_\perp$ where $\Sigma_i$ no longer decreases monotonically as $b_\perp$ increases (typically because the profile has run into another cloud or high-extinction structure in the interstellar medium, at least in projection). The average in Equation \ref{eq:colavg} extends parallel to the spine/symmetry axis until all bins have exited the $A_V=1$ contour.

This procedure is repeated $N_\mathrm{boot}$ times with slightly different choices in the parameters that define the procedure (namely $A_{V,\mathrm{high}}$ and the adopted distance), and with maps produced by sampling pixels from the entire original map with replacement such that the number of pixels in each of the resampled maps is equal to the number of pixels in the original map. $A_{V,\mathrm{high}}$ is varied uniformly between the $97.5$th and $99.5$th percentile, and distance is varied by 10\% given the observed scatter between the dust-based and VLBI-based \citep[e.g.][]{Reid_2014,Reid_2016, Brunthaler_2011, Loinard_2013} distances \citep{Zucker_2020}. This procedure produces $N_\mathrm{boot}$ replicates of $\Sigma_i$, allowing us to estimate the covariance of the $\Sigma_i$ via the sample covariance:
\begin{equation}
    \hat{S}_{ij} = \frac{1}{N_\mathrm{boot}} \sum_{k=1}^{N_\mathrm{boot}}(\Sigma_{i,k} - \bar{\Sigma}_i)(\Sigma_{j,k} - \bar{\Sigma}_j),
\end{equation}
where $\Sigma_{i,k}$ denotes the $k$th replicate of $\Sigma_i$ and $\bar{\Sigma}_i$ is the mean value of $\Sigma_i$ across the $N_\mathrm{boot}$ replicates. In general, neighboring $\Sigma_i$ may be correlated or anti-correlated because of variations in the adopted distance and the exact position and orientation of the spine. By resampling pixels from the entire map, we produce a robust estimate of the $\Sigma_i$, their uncertainties, and their covariance across radial bins regardless of the underlying, non-Gaussian, distribution of the pixels that enter the average that determines $\Sigma_i$ (see Equation \ref{eq:colavg}) including upstream effects like where exactly the spine is placed.

\begin{figure*}
\begin{centering}
\includegraphics[width=\textwidth]{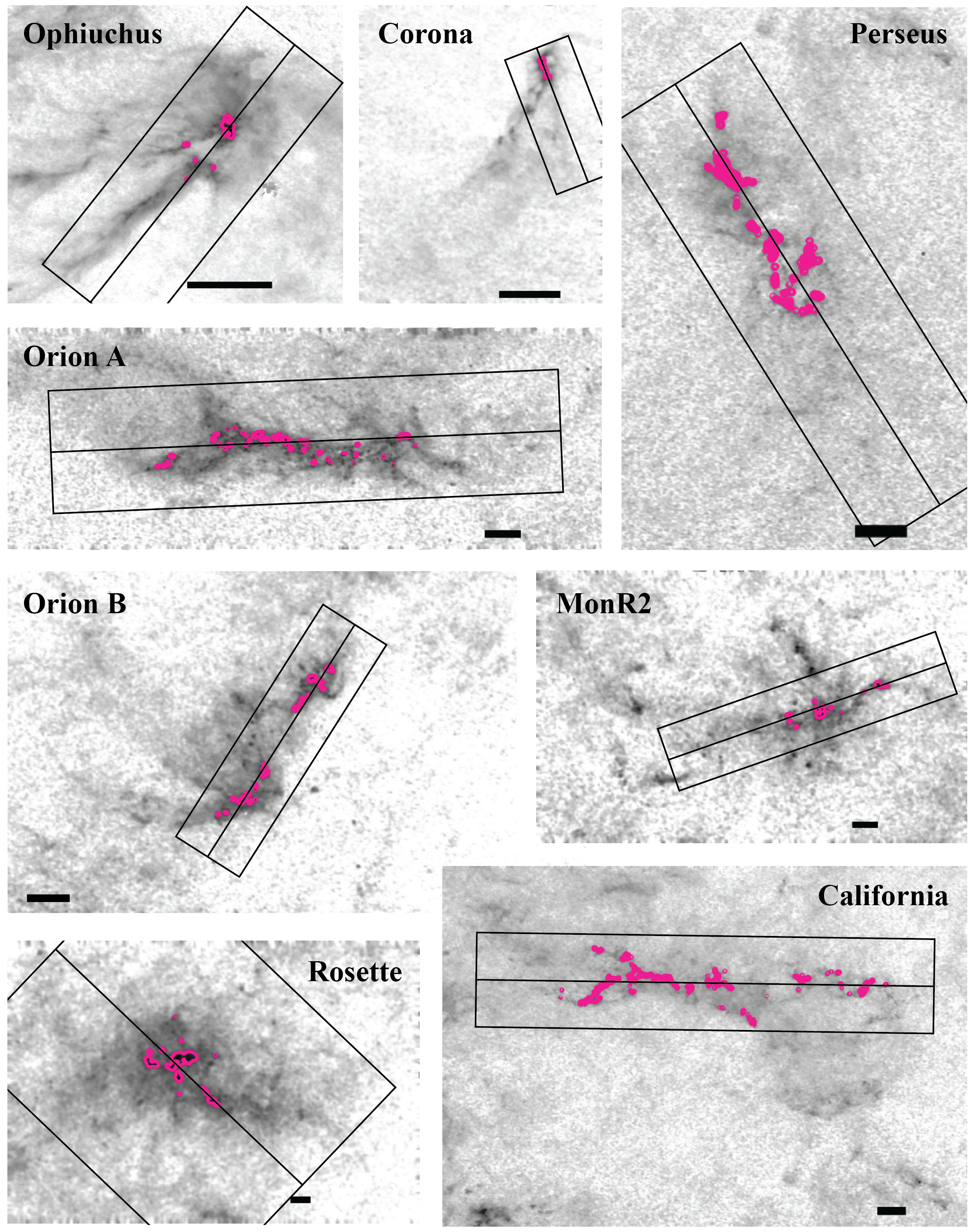}
\caption{Extinction maps for the clouds in our sample, as defined in Table 1. Each black box represents the boundary used to compute the cloud's column density profile and mass. The central line shows a typical spine, defined using the high-extinction contours (pink). The black scale bar represents $5$ pc at the adopted median distance to the cloud.}
    \label{fig:mosaic}
\end{centering}
\end{figure*}

In addition to the $\Sigma_i$, we can also compute any quantity that depends on the map and/or the cloud's distance, along with estimates of its covariances with any other such quantity. Of particular importance is each cloud's mass, either including or excluding thresholds in $A_V$. Masses are computed as
\begin{equation}
    M = \sum_j \mathcal{C} A_{V,j} d^2 \Delta l \Delta b \mathcal{I}_j,
\end{equation}
where $d$ is the distance to the cloud, $\Delta l$ and $\Delta b$ are the (equal) pixel scales in each direction of the map (in radians), and $\mathcal{I}_j$ is 1 if and only if the following conditions are met:
\begin{itemize}
    \item The pixel has a $b_\perp < b_{\perp,\max}$.
    \item At the pixel's coordinate along the spine of the cylinder ($b_\parallel$), there must be at least one $b_\perp$ bin within the $A_V=1$ contour. That is, the length (parallel to the spine) of the rectangular area on the sky defining the cloud does not stretch beyond the $A_V=1$ contour.
    \item The pixel is within the $l$ and $b$ bounds defined in Table 1.
\end{itemize}
These conditions correspond to the boundaries shown in Figure \ref{fig:mosaic}.

Given an estimate of $S_{ij}$, we can now reasonably evaluate the likelihood of an observed surface density profile $\Sigma_i$ given fixed values of the model parameters $\rho_0$, $\alpha$, $r_0$, and $R_T$. To account for additional systematic uncertainties, including the deliberately simple nature of the model, we also include a parameter $\sigma^2$ that is added to each diagonal component of $\hat{S}$. We note that this simple addition does not naturally account for unmodelled correlated errors, but is the simplest and most interpretable addition to the model of this sort. We also emphasize that we do explicitly account for the most prominent source of correlated errors, namely uncertainty in the distance, in our bootstrapping procedure. The likelihood is therefore a multivariate normal:

\begin{equation}\label{eq:likelihood}
\begin{split}
    \mathcal{L} &= p(\{\Sigma_i\}|\rho_0/\cos~i,\alpha,r_0,R_T,\sigma^2) \\
    &= \frac{1}{\sqrt{(2\pi)^{N} \det{\mathcal{S}}}}\exp\left(-\frac{1}{2} z^T \mathcal{S}^{-1} z \right),
\end{split}
\end{equation}
where $z$ denotes the displacement vector with components

\begin{equation}
    z_i = \Sigma_i - \Sigma_\mathrm{model}(b_{\perp, i}; \rho_0/\cos~i,\alpha,r_0,R_T)
\end{equation}
$\Sigma_\mathrm{model}$ is the same quantity defined in Equation \ref{eq:surf}
at the center of the bin $b_{\perp,i}$. The covariance matrix has components

\begin{equation}
\mathcal{S}_{ij} = \hat{S}_{ij} + \sigma^2\delta_{ij},
\end{equation}
where
\begin{equation}\label{eq:pinf}
\delta_{ij} = 
\begin{cases}
1    &   \text{if}~ i=j, \\
0    &   \text{if}~i\ne j.
\end{cases}
\end{equation}
\noindent
The $N$ appearing in Equation \ref{eq:likelihood} is the number of bins of $b_\perp$, which is also the size of the $z$ vectors and the size in each dimension of $\mathcal{S}$.

\input{table_prior}

To fit this model we need to specify a prior distribution of the model parameters. We choose broad/non-informative priors on the log of each variable independently. These distributions are listed in Table 2. While the boundaries of these distributions are not generally important, given that the data are able to constrain each parameter to at least some degree, we do impose an upper limit on $\log \sigma^2$ equal to the log of the variance of the $\Sigma_i$, minus an additional (0.6 dex)$^2$, to remove a branch of the posterior distribution in which the physical parameters $\rho_0/\cos~i$, $\alpha$, $r_0$, and $R_T$ play no role, and all variation in the data is attributed to scatter from $\sigma^2$. The value of 0.6 is likely related to the typical number of effective radial samples for our clouds, but was chosen in practice by running the fits with a larger upper limit and noting the upper limit of $\log \sigma^2$ necessary to remove the unphysical part of the posterior. Once this branch is removed, $\sigma^2$ has little additional effect on the inferred parameters for most clouds (see Appendix). Samples from the posterior distribution are drawn with the Python package \texttt{emcee} \citep{Goodman_2010, ForemanMackey_2013}. The results of this analysis and the corresponding fits are shown in the next section.

\section{Results}\label{sec:results}

\begin{figure*}[h!]
\begin{centering}
\includegraphics[width=0.8\textwidth]{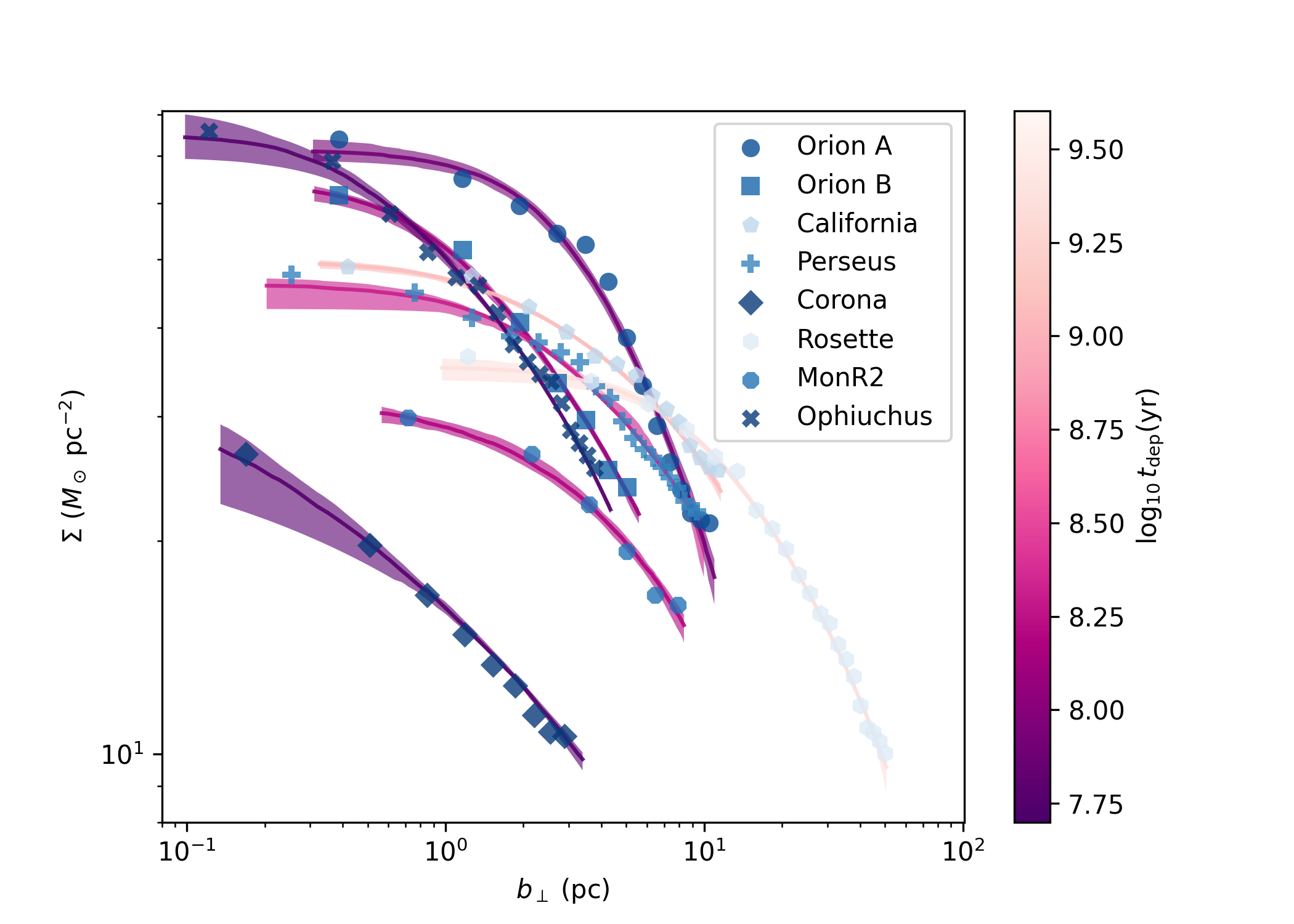}
\caption{Observed radially averaged surface densities (blue circles) as a function of $b_\perp$.  Pink shaded areas (pink lines) show the central 68\% confidence interval (median) for the conditional distribution $\Sigma | b_\perp$ over a finely-sampled range of values for $b_\perp$.  \label{fig:profiles}}
\end{centering}
\end{figure*}

\begin{figure*}
\includegraphics[width=\textwidth]{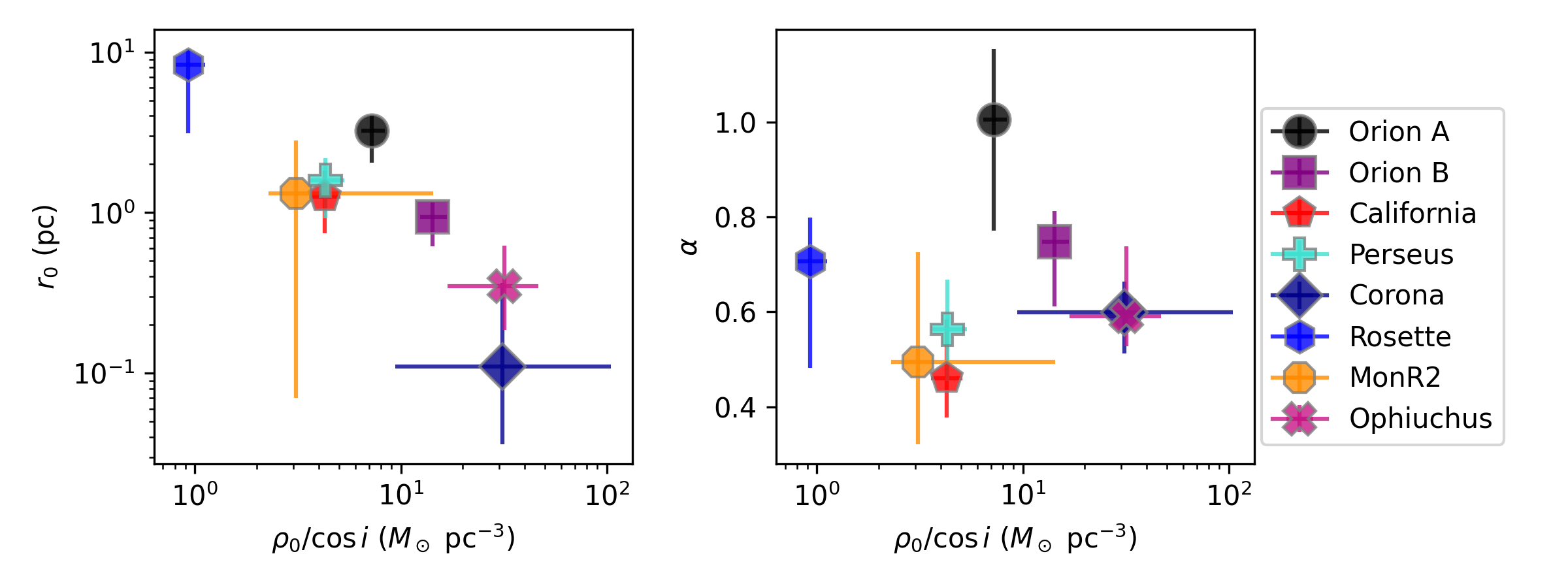}
\caption{Results of fitting the explicit cylinder model. Each cloud's median posterior value of $\rho_0$, $r_0$, and $\alpha$ is plotted, with errorbars showing the central 68\% of the marginal distribution of each parameter. Clouds with higher central densities $\rho_0$ tend to have more compact centers (lower $r_0$), but all clouds have $\alpha$ between about 0.5 and 1. \label{fig:r0rho0} }
\end{figure*}

\begin{figure*}
\includegraphics[width=\textwidth]{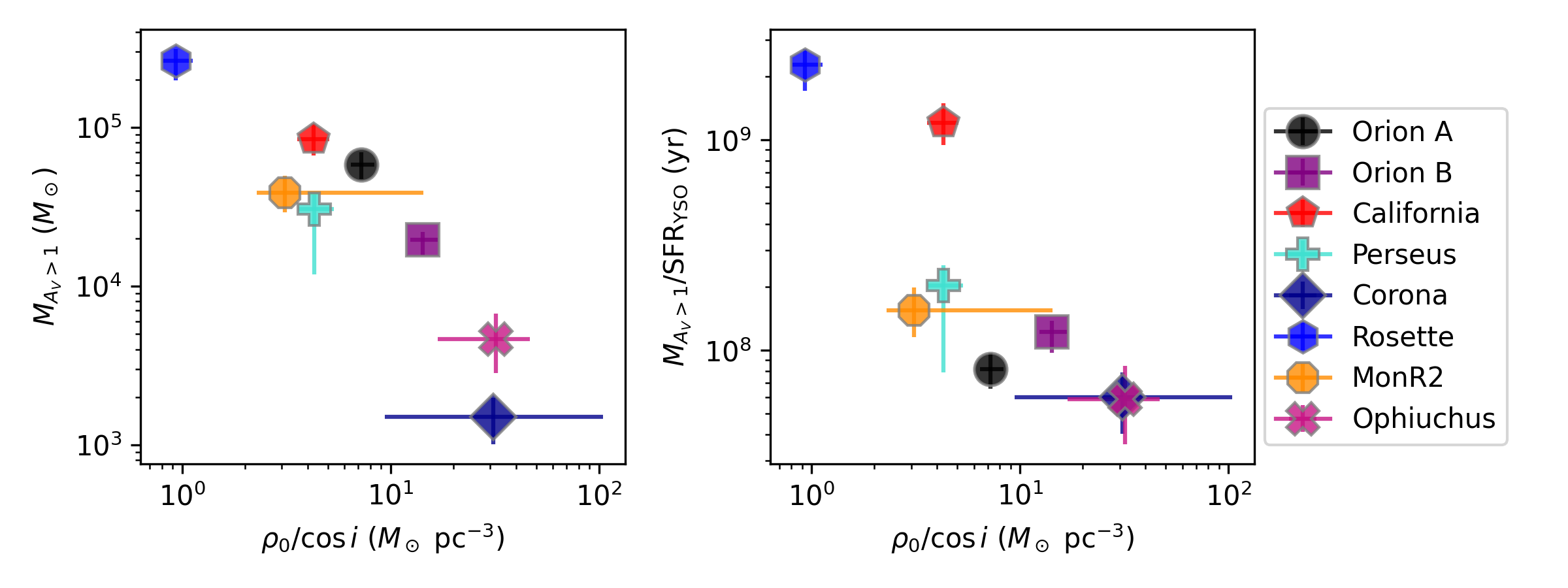}
\caption{The central density of each cloud relative to its mass and depletion time. Clouds with higher central densities tend to have lower masses and shorter depletion times. \label{fig:rho0mass} }
\end{figure*}

Figure \ref{fig:profiles} displays the averaged radial dust column density profiles of the eight clouds in our sample. The high resolution of the data allow a careful comparison with our model. We see that for each cloud there is a solution that fits the data remarkably well. The blue points show the $\Sigma_i$ for each cloud, while the pink shaded areas (pink lines) show the central 68\% confidence interval (median) for the conditional distribution $\Sigma | b_\perp$ over a finely-sampled range of values for $b_\perp$. Generally the model is a good description of the data points, with uncertainties increasing towards $b_\perp \rightarrow 0$, given the finite resolution of the maps and hence the binned $\Sigma_i$. The same posterior distribution represented in Figure \ref{fig:profiles} is shown in a variety of ways in the rest of this section.

Figure \ref{fig:r0rho0} shows the relationships between the main model parameters, $\rho_0/\cos~i$, $r_0$, and $\alpha$. There is a clear anti-correlation between $\rho_0/\cos~i$ and $r_0$, such that clouds with higher central volume densities are more ``compact'', at least inasfar as they have smaller scale radii $r_0$. Meanwhile $\alpha$, which determines the fall-off of density with radius for $r\gg r_0$ (see Equation \ref{eq:rho1}), has no apparent relationship with $\rho_0/\cos~i$ or $r_0$, simply falling within a range of $\sim 0.5-1$, broadly similar to previous estimates of filamentary density profiles \citep[e.g.,][]{Arzoumanian_2011, Palmeirim_2013, Toci_2015, Zucker_2018, Arzoumanian_2019}.

Concentrating on $\rho_0/\cos~i$, which has a strong relationship with $r_0$, we compare $\rho_0$ to each cloud's mass within its $A_V>1$ contour and the cloud's corresponding depletion time, defined as its mass divided by its star formation rate (Figure \ref{fig:rho0mass}). The clouds' SFRs are derived from counts of their young stellar objects (YSOs) based on infrared data \citep[e.g.,][]{Hillenbrand_1997, Lada_2009}, and  applying the conversion
\begin{equation}
    \mathrm{SFR} = N_\mathrm{YSO} \cdot 0.25 \cdot 10^{-6} M_\odot\ \mathrm{yr}^{-1},
\end{equation}
from \citet{Lada_2010}. See Table 2 of \citet{Lada_2010} for additional references on YSO counts for most clouds, and \citet{Gutermuth_2011} for MonR2.

We find that clouds with the highest central volume densities in our sample tend to have the lowest masses and the shortest depletion times, or equivalently the highest SFRs relative to their masses. It is to be expected that star formation might proceed more vigorously in denser gas given that the free-fall time scales as $\rho^{-1/2}$, but it is less intuitive that higher central volume densities should correspond to lower masses. We speculate on the ultimate cause of this correlation in the next section.

\begin{figure*}
\begin{centering}
\includegraphics[width=0.8\textwidth]{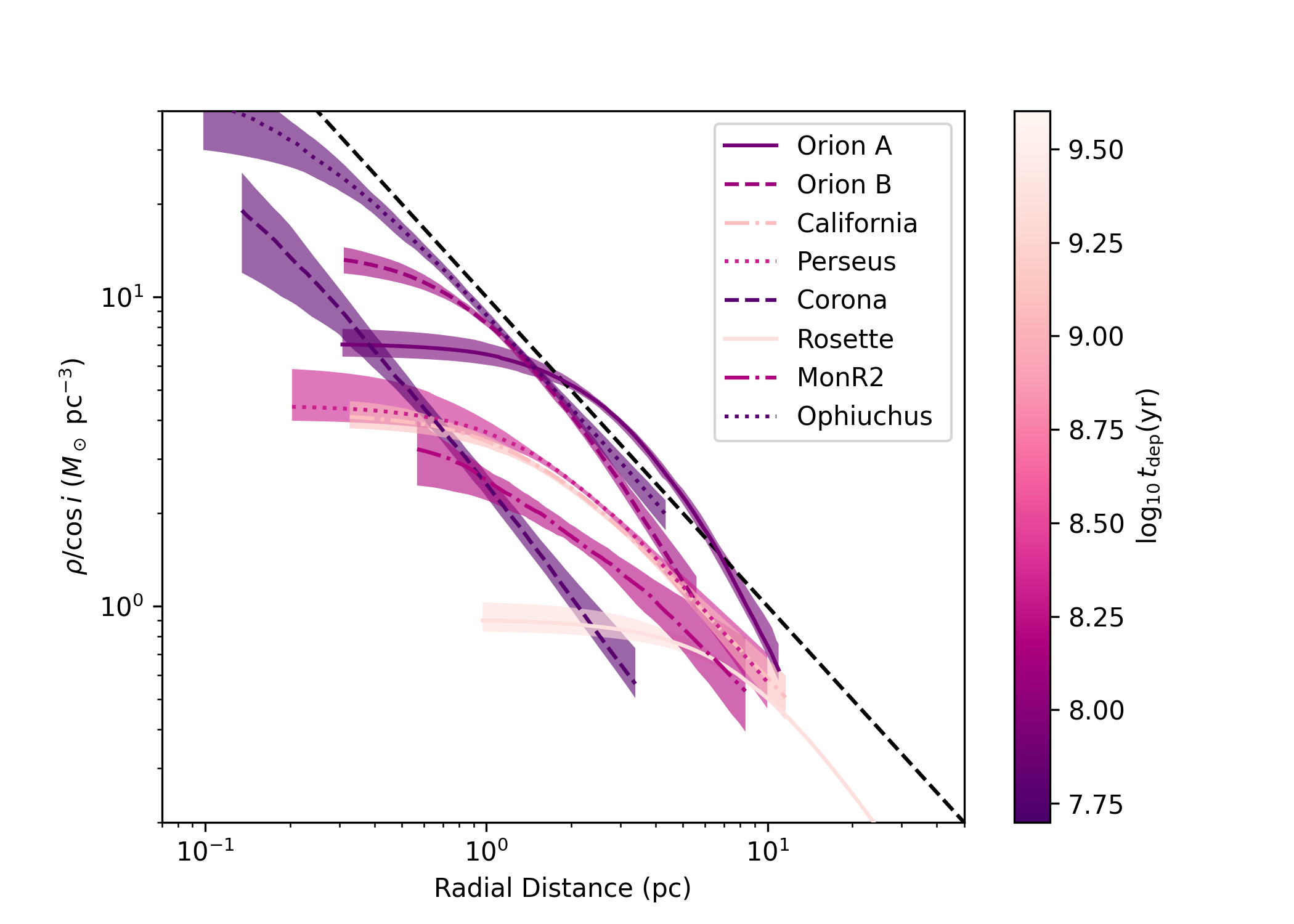}
\caption{Inferred density profiles as a function of radial distance. As in Figure \ref{fig:profiles}, the conditional posterior distribution at each radius is represented by a line at the median value for each cloud and a shaded region showing the central 68\% of the distribution. Instead of $\Sigma|b_\perp$, here we show the distribution of $\rho|r$ for each cloud. The black dashed line shows $\rho_\mathrm{env} = 10(r/\mathrm{pc})^{-1}\ M_\odot\ \mathrm{pc}^{-3}$.} \label{fig:densprofiles}
\end{centering}
\end{figure*}

\section{Discussion}\label{sec:discussion}
In fitting our explicit but simple model of molecular clouds as cylinders, we have come across several interesting correlations. The clouds' central volume densities $\rho_0/\cos~i$ seem to be closely tied to their scale radius $r_0$, their mass $M_{A_V>1}$, and their depletion times, $M_{A_V>1}/\mathrm{SFR}$, but with little correspondence to their outer powerlaw slopes $-2\alpha$, or their truncation radius $R_T$. To delve into these potential relationships, we have found it helpful to plot the inferred density profiles as a function of radial distance from the spine in Figure \ref{fig:densprofiles}. 

While there is still a fair amount of variety between the clouds, some of which may come from the unknown inclination $i$, we see that roughly speaking the clouds' density profiles fit within an envelope of $\rho_\mathrm{env} \approx 10\ (r/\mathrm{pc})^{-1} M_\odot\ \mathrm{pc}^{-3}$, with each cloud having a different radius and density at which it turns over and flattens towards $r\rightarrow 0$. In this sense the relationship between $r_0$ and $\rho_0/\cos~i$ becomes clear: the higher the central density, the smaller the turnover radius must be in order to remain within the $\rho/\cos~i \lesssim 1/r$ envelope. Indeed $r_0$ should be roughly $10~\mathrm{pc} / (\rho_0/\cos i / M_\odot\ \mathrm{pc}^{-3})$, exactly as shown in Figure \ref{fig:r0rho0}.

The interpretation of the clouds' density profiles as lying within this envelope also helps explain the clouds' masses, and in particular the somewhat surprising fact that clouds with higher central densities have lower total masses. The differential element of mass for a fixed apparent length\footnote{The true length of the cloud is $L = b_{\parallel,\max}/\cos~i$.} of the cloud $b_{\parallel,\max}$ is $2\pi r b_{\parallel,\max} \rho(r)/\cos~i ~dr$; thus, as long as $\rho(r)$ is less steep than $1/r$ near the cloud center, the exact density profile in the central pc will not have a large effect on the cloud's mass. Instead, the clouds' masses will be dominated by the large-radius behavior, for which we can use the $1/r$ envelope, obtaining
\begin{equation}
M \sim \int 2\pi r b_{\parallel,\max} \rho_\mathrm{env} dr \approx 70~ M_\odot \left(\frac{b_{\parallel,\max}}{\mathrm{pc}}\right) \left(\frac{b_{\perp,\max}}{\mathrm{pc}}\right).
\end{equation}
This is of course equivalent to stating that within the boundary of the cloud---in this case a rectangular area on the sky of size $b_{\parallel,\max}$ by $b_{\perp,\max}$---the average surface density is $\sim 70~M_\odot\ \mathrm{pc}^{-2}$, similar to Larson's classic result \citep{Larson_1981} and more recent results on molecular filaments \citep{Zhang_2019}. In this framing the question then becomes: how do $b_{\parallel,\max}$ and $b_{\perp,\max}$ scale? 

\begin{figure*}
\includegraphics[width=\textwidth]{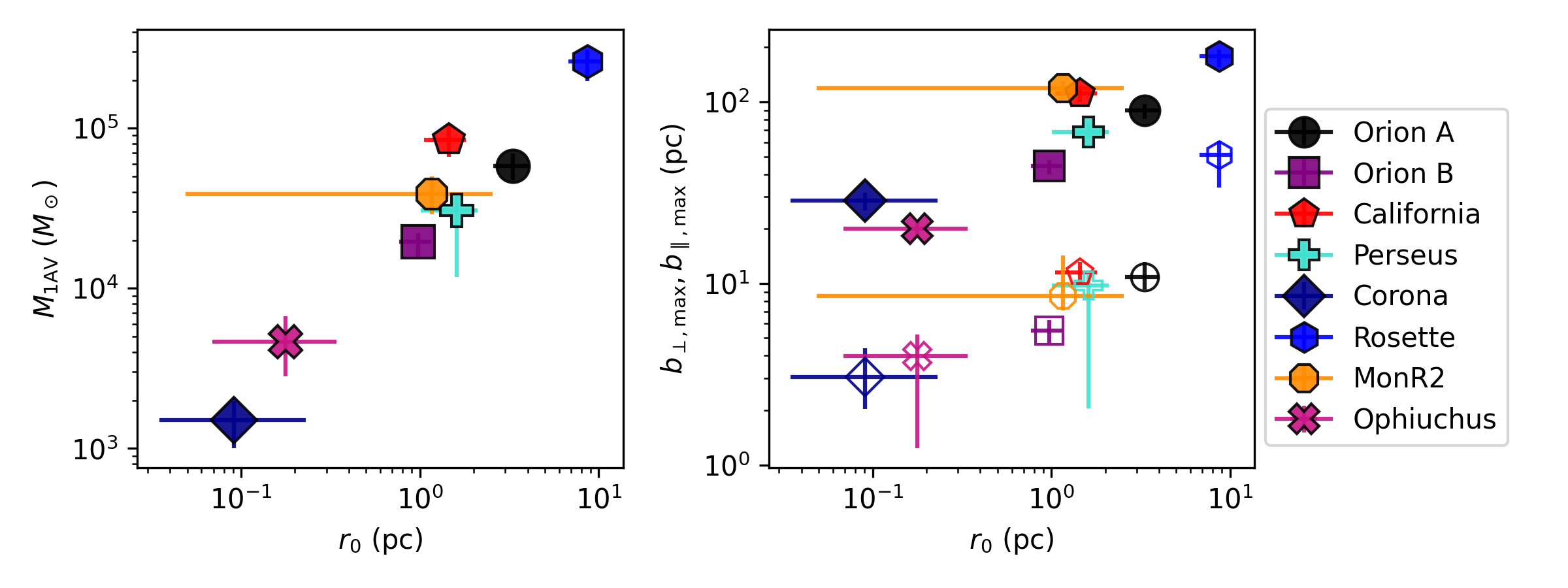}
\caption{The scaling of cloud mass (left panel) and $b_{\parallel,\max}$ and $b_{\perp,\max}$ (right panel) with $r_0$. In the right panel, the $b_{\perp,\max}$ have smaller values, and the markers are not filled. $b_{\parallel,\max}$ and $b_{\perp,\max}$ both scale with $r_0$ but with a logarithmic slope less than 1.} \label{fig:rscaling} 
\end{figure*}

Since $M$ scales roughly as $r_0$ (see left panel of Figure \ref{fig:rscaling}), it is the product $b_{\parallel,\max} b_{\perp,\max}$, not either length scale individually, that scales as $r_0$. We find that $b_{\parallel,\max}$ and $b_{\perp,\max}$ both have a modest trend with $r_0$, with each contributing about equally to the scaling of the area $b_{\parallel,\max} b_{\perp,\max}$ (see right panel of Figure \ref{fig:rscaling}). By splitting the ``size'' of the cloud into two separate dimensions, we recover a more modest relationship between cloud size and mass than in \citet{Larson_1981}.

The star formation rate, meanwhile, is expected to obey some relationship with the volume density of the gas \citep{Schmidt_1959, Kennicutt_2012}. For volume densities, the density of the star formation rate is commonly expressed as
\begin{equation}
\label{eq:ks}
    \dot{\rho}_\mathrm{SFR} = \epsilon_\mathrm{ff} \frac{\rho}{t_\mathrm{ff}},
\end{equation}
where $\epsilon_\mathrm{ff}$ is an efficiency factor of order 1\% \citep[e.g.][]{krumholz_2007, krumholz_2012}, and the free-fall time $t_\mathrm{ff} = \sqrt{3\pi/32G\rho}$. Interestingly, integrating Equation \ref{eq:ks} over the volume of the model cylinders does not reproduce the observed SFRs as measured by YSO counts, except for California and Rosette. In most other clouds Equation \ref{eq:ks} applied to the smooth model falls short of the actual SFR by about a factor of 10. The unknown inclination of the clouds lowers the expected star formation rate from integrating Equation \ref{eq:ks} by an additional factor\footnote{The average value of $\cos~i$ assuming randomly-oriented clouds is $\pi/4\approx 0.79$, and the average value of $(\cos~i)^{1.5}$ is about $0.72$, so typically the inclination effect will be non-negligible but still only $\sim30\%$.} of $(\cos~i)^{1.5}$.  Given this gap and the variation in its size across clouds, when contrasted with the simple approximately powerlaw relationship between $\rho_0/\cos~i$ and $t_\mathrm{dep}$, the SFR cannot be explained as an integral of our smooth global density distribution. We speculate that if we had access to the true 3D density distribution, integrating Equation \ref{eq:ks} would recover something much closer to the true SFR, but because our smoothed distribution averages out high-density peaks, we under-estimate the SFR by integrating the analytic model.

While the inferred large-scale density distribution of the cloud cannot be used to infer the SFR directly by integration of Equation \ref{eq:ks}, there is still a trend between the depletion time and the parameters of the large-scale cloud model. Very roughly $t_\mathrm{dep} \propto (\rho_0/\cos~i)^{-1}$. Gas governed by Equation \ref{eq:ks} has a depletion time $t_\mathrm{ff}/\epsilon_\mathrm{ff} \propto \rho^{-1/2}$. This discrepancy in powerlaw index is manifestly not explained by suitable averaging of $\rho$ over the volume of the cylinder (see previous paragraph). This is consistent with the fact that such an integral is dominated by the region where $\rho \approx \rho_0$, which would produce a $t_\mathrm{dep}$ that scales as $\rho_0^{-1/2}$. We speculate that the steeper relationship observed in these clouds is the result of the global evolutionary state of the clouds, where clouds with higher $\rho_0/\cos~i$ are further along in their evolution and hence have more observable YSOs, a possibility we will explore in more detail in future work.

\begin{figure*}
\includegraphics[width=\textwidth]{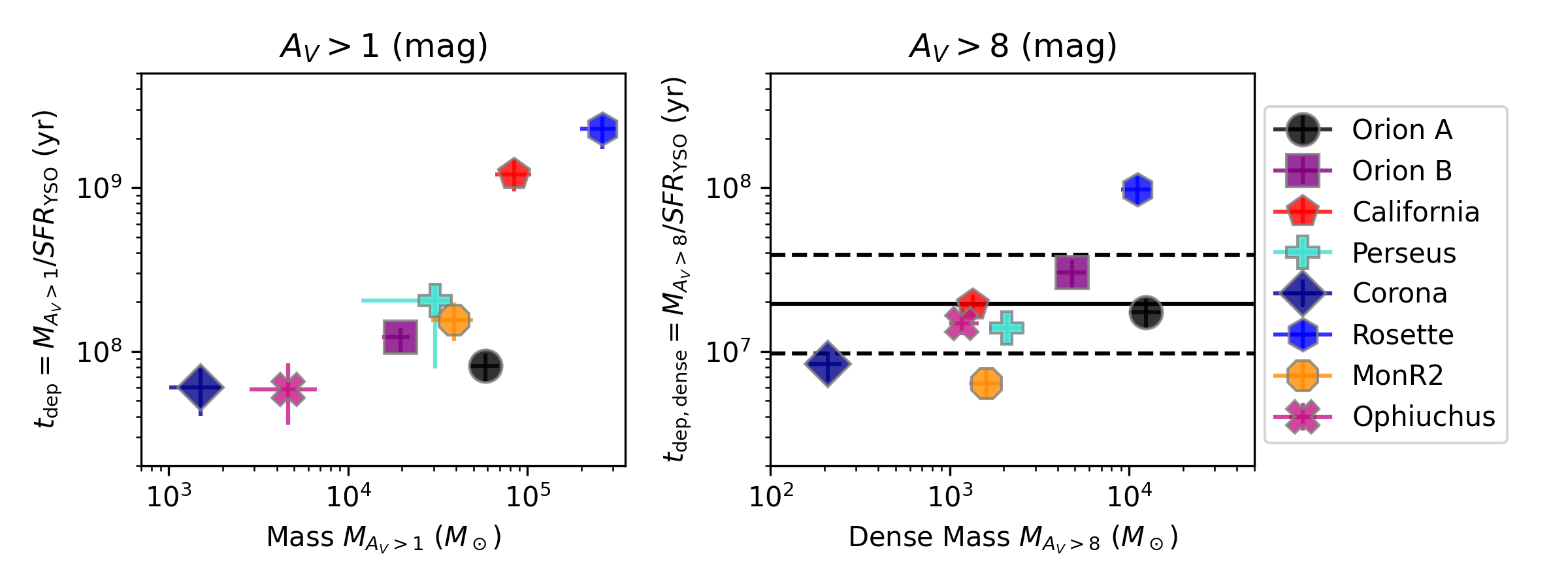}
\caption{Depletion time as a function of mass, as defined for material within $\av > 1$ mag (left) and $\av > 8$ mag (right).  Dashed lines in the right panel show a factor of $\pm 2$ spread (from the solid line) in dense gas depletion time, comparable to the extremes of the \cite{Lada_2010} result. Note that both panels have the same dynamic range on the y-axis. While we reproduce the \citet{Lada_2010} result that spread in depletion times becomes smaller when focusing on gas in high-extinction regions, we do see a trend in both panels that higher-mass clouds have longer depletion times.  \label{fig:vslada}}
\end{figure*}

\citet{Lada_2010}, hereafter LLA10, showed using the same dust extinction data and a similar sample of local molecular clouds that depletion time has no discernible trend with cloud mass, defined as mass within the $A_V=1$ contour, and that the spread in depletion times becomes much smaller when using the mass within a contour of $A_V\ga 8$. 

In Figure \ref{fig:vslada} we show the trends between mass and depletion time for both $A_V>1$ and $A_V>8$. The two panels in the plot have the same dynamic range of depletion times on the y-axis, though offset in absolute value by a factor of 10, since the dense mass (mass with $A_V>8$) of a cloud is less than the total mass (mass with $A_V>1$). The horizontal black lines in the right panel show a factor of two spread in dense gas depletion time, comparable to the extremes of the LLA10 result. Where LLA10 found no trend with total mass, we find that higher-mass clouds have longer depletion times, regardless of whether we look at total mass or dense mass. We do recover the main result that the scatter in depletion times is substantially reduced when looking at the dense mass, though not quite as much as in LLA10. 

We can attribute these differences to four factors: first, we have used updated distances to the clouds based on Gaia distances \citep{Zucker_2020}; this has the largest effect on the mass estimates of the clouds when compared with estimates from LLA10 and \citet{Lombardi_2011}. Second, we have used a somewhat different selection of clouds, for instance excluding a number of clouds that LLA10 included with low masses and long depletion times (Pipe, and Lupus 1, 3, and 4), while including clouds excluded by LLA10 (Rosette, MonR2). The former set of clouds have substantial uncertainties associated with their distances and, for Pipe, with subtracting background dust absorption in the extinction map. Third, we have used slightly different definitions of the clouds and their masses, though these differences are small compared to the updates to the clouds' distances. Fourth, our method produces confidence intervals for the cloud masses (shown with errorbars in Figure \ref{fig:vslada} and throughout this paper), which changes the assessment of which trends are plausible vs implausible.

While our analysis largely confirms the main result of LLA10, that the spread in depletion times is reduced substantially by using only the high-extinction gas, we believe this can only be a partial explanation of the origin of the SFR in these clouds. High-extinction areas in a 2D map can only be proxies the 3D configuration of gas in a cloud. Moreover there appear to be residual trends between cloud mass and depletion time, even upon restricting to the high-extinction gas, suggesting that the surface density of the gas at the present moment is not the last word on the star formation law. Our model for the large-scale density structure of the clouds averages out small-scale density fluctuations, yet produces correlations with the observed depletion times, suggesting that what was missing from LLA10 is some accounting for the large-scale structure of each clouds as we have provided here.

While our results provide an interesting new perspective on well-studied local molecular clouds, they also have broader implications. Models in a variety of contexts depend explicitly or implicitly on breaking the star formation rate in a patch of the interstellar medium, a whole galaxy, or even across galaxies, into a population of molecular clouds. The creation, destruction, and time-dependent star formation within each molecular cloud contributes to variability in the integrated star formation rate \citep{Tacchella_2020, Chevance_2020}. This variability has been invoked to explain the discovery by JWST of galaxies at $z\sim 6$ that have, at least momentarily, stopped forming stars \citep{Dome_2023}. Extragalactic surveys with large sample sizes are able to constrain the cloud scaling relations that enter into these models \citep[e.g.][]{Hughes_2013, Faesi_2018, Rosolowsky_2021}, but the more-limited spatial resolution and sensitivity make morphological study beyond near-spherical symmetry more difficult. For instance, \citet{Hughes_2013} recover axis ratios around 2 for their population of clouds in M51, M33, and the LMC, as compared to typical values for the local clouds in our sample of around 10 (see Figure \ref{fig:rscaling}).

Just as the global star formation rate can be broken into its contributions from individual clouds, the line luminosities from star-forming regions at high redshift are summed from individual clouds \citep{Narayanan_2014, Krumholz_2014, Narayanan_2017}. The size and morphology explicitly enters these models, which are then summed to predict line luminosities as a function of galaxy properties \citep{ Popping_2019, Yang_2022}. Changes in the underlying assumptions about the typical morphology of clouds, i.e. from spherical to cylindrical or filamentary, may alter the inference of the galaxy property distribution \citep{Zhang_2023} from forthcoming emission line intensity mapping surveys \citep[e.g.][]{Pullen_2023}.

\section{Summary}\label{sec:summary}

We proposed a simple 3D model of global molecular cloud morphology and tested it using dust extinction measurements of a sample of Solar Neighborhood clouds having a range of SFRs. We find that for each cloud there is a model solution that fits the data extremely well. Our most striking results are as follows:

\begin{enumerate}
    \item There is a clear anti-correlation between the central volume density $\rho_0$ and the turnover radius $r_0$, implying that clouds with higher $\rho_0$ are more compact. This can be explained by observing that the clouds' density profiles fit within an envelope of $\rho_\mathrm{env} \approx 10~r^{-1} M_\odot\ \mathrm{pc}^{-3}$. The higher the central density, the smaller $r_0$ must be in order to remain within the $\rho \lesssim 1/r$ envelope.

    \item Somewhat surprisingly, clouds with higher central densities have lower total masses. This is because the density profile in the central pc does not have a large effect on a cloud’s mass, as long as $\rho(r)$ is less steep than $1/r$ near the cloud center. Clouds’ masses will be dominated by the large-radius behavior, described by the $1/r$ envelope.
    
    \item Higher-mass clouds have longer depletion times, regardless of whether we consider their total mass or dense mass. 

    \item While the overall dense gas fraction of molecular clouds helps to explain some of the variation in the depletion time, our results show that the global cylindrical structure of clouds---reflected for instance in the central volume density---plays a key role in determining $t_{\rm dep}$. 

\end{enumerate}

\acknowledgements
We thank the anonymous referee for a thoughtful and helpful report. JCF thanks the Simons Foundation for support via a Flatiron Research Fellowship.


\appendix

\begin{figure*}
\includegraphics[width=0.9\textwidth]{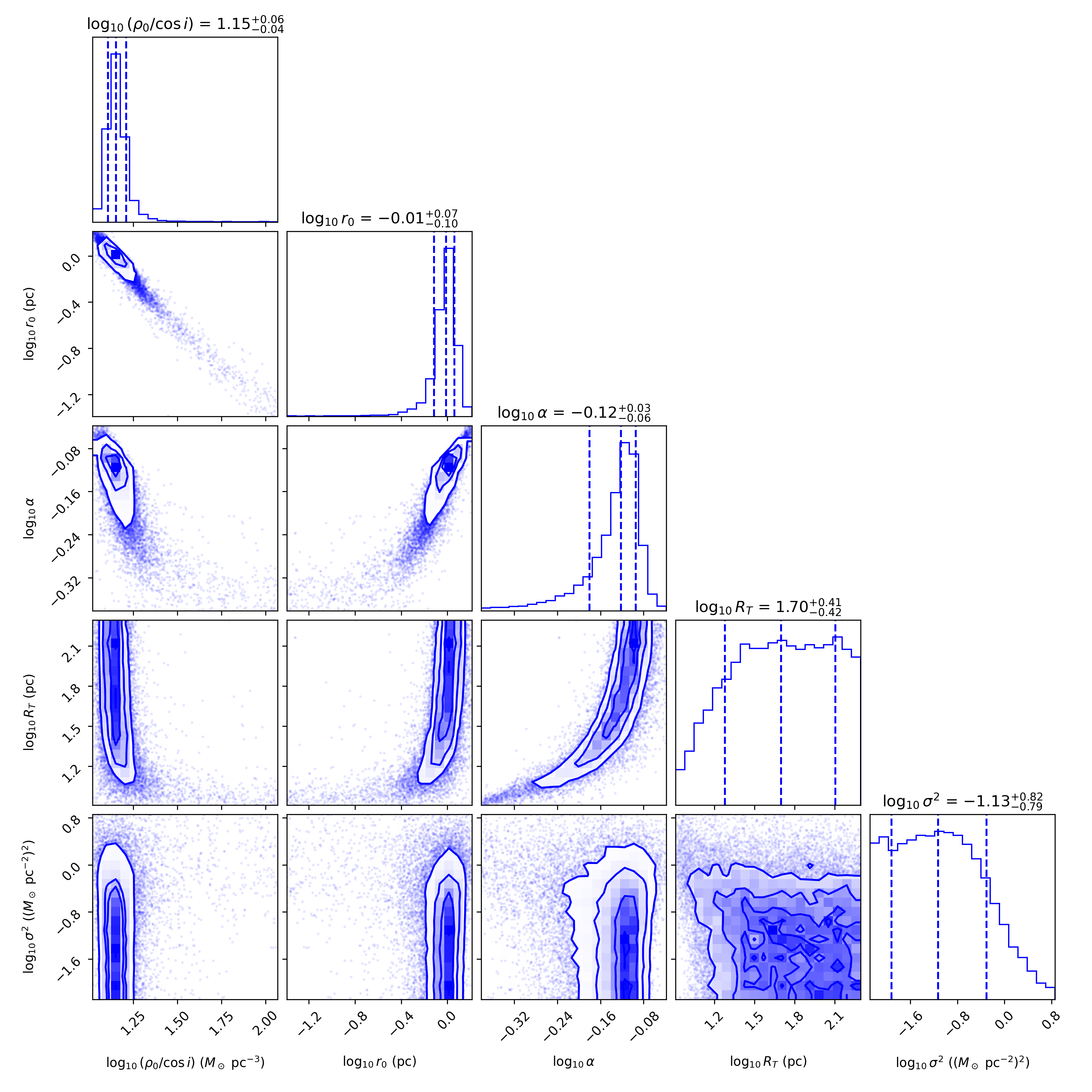}
\caption{A corner plot for our model fit to the Orion B molecular cloud. The histograms on the diagonal show the 1D marginal posterior distributions of the 5 parameters in the model, while the off-diagonal plots show each 2D marginal posterior distribution, representing the distribution through a combination of contours and individual points in less sparsely-populated areas.  \label{fig:corner}}
\end{figure*}

A key step in our analysis is the fitting of an explicit parameterized model for the 3D density structure of molecular clouds to the column density profiles of each molecular cloud in our sample. Throughout this work we typically show points with errorbars for posterior predictive quantities, where the points show the median of the relevant 1D marginal posterior distribution, and the errorbars show the central 68\% of the distribution. However, it is often helpful to look at a more systematic, though still low-dimensional, representation of the posterior distribution to understand nontrivial features of the posterior distribution like degeneracies, sensitivity to the prior distribution, and multimodality. We therefore show an example corner plot \citep{corner}, in this case for our fit to the Orion B molecular cloud, and explain its features in some detail. The features present in this corner plot are similar though of course not identical for the other clouds in the sample.

The summary of the fit shown in Figure \ref{fig:corner} displays several interesting features, including both linear and non-linear degeneracies, and several upper/lower limits. It is worth discussing physically where these features come from and their effects on the results presented in this work. The most important is probably the 1:1 degeneracy between $\log_{10} r_0$ and $\log_{10} (\rho_0/\cos~i)$. This is potentially concerning since one of our main conclusions is that the clouds display an anti-correlation between $r_0$ and $\rho_0$ -- if all clouds displayed a strong degeneracy between these two parameters, the best-fit $\rho_0$ and $r_0$'s of a population of clouds may appear to be anti-correlated even if no such relationship existed intrinsically. It is therefore worth explaining where this degeneracy in the model comes from and how concerned we should be about it.

In the limit where $r\gg r_0$, the density profile according to Equation \ref{eq:rho1} is just $\rho_0 (r_0/r)^{2\alpha}$. It follows that at fixed $\alpha$, the model would predict identical density profiles for any value of $r_0$ so long as $\rho_0 r_0^{2\alpha}$ is held constant. Despite this degeneracy, however, Orion B's values of $\rho_0$ and $r_0$ are quite tightly constrained, with only a small fraction of samples extending to large values of $\rho_0/\cos~i$. This is because Orion B, like most clouds in the sample, includes constraints from $b_\perp < r_0$, which breaks the degeneracy between $r_0$ and $\rho_0$. A similar qualitative result holds for most other clouds with the exception of Corona, where all of the datapoints are outside $r_0$. For Corona, $r_0$ and $\rho_0$ are more poorly-constrained, with $\rho_0/\cos~i$ extending up to the maximum allowed prior value of $10^{2.3}$. In any case, the constraints on $\rho_0$ and $r_0$ are individually somewhat worse because of this degeneracy, but this is included in the size of the errorbars.

Figure \ref{fig:corner} also makes it clear that the constraint on $R_T$ is a lower limit. Once $R_T$ is much larger than the values of $b_\perp$ in the profile, its exact value becomes unimportant since the line of sight integrals (Equation \ref{eq:surf}) are dominated by the highest-density part of the integral, that is the part of the line of sight closest in radial distance to the spine, so long as $\alpha$ is sufficiently large. When $\alpha$ is smaller, $R_T$ does matter, as can be seen in the 2D marginal distribution between $\log_{10}\alpha$ and $\log_{10} R_T$. 

Meanwhile, the fit shows that $\log_{10}\sigma^2$ is apparently best understood as having an upper limit around $0.5 \mathrm{dex}$. Values of $\log_{10}\sigma^2$ greater than about $-0.8$ begin to not allow the data to fit any better while widening the peak of the likelihood distribution. For Orion B in particular, the exact value of $\sigma^2$ becomes unimportant when $\sigma^2$ is small because the model on its own is sufficiently flexible to fit the data well using the already-modelled errors estimated from the bootstrapping procedure. Eventually if we allowed $\sigma^2$ to increase even further, another branch of the posterior distribution would appear in which all of the variation in the data would be explained by $\sigma^2$, and none would be explained by the model itself. This would appear in the 2D marginal distributions with $\sigma^2$ as regions where $\sigma^2$ would be large, and every other parameter would have a flat distribution in $\log$ space (the prior). By limiting $\log_{10}\sigma^2$ to the variance in the data points, minus an additional $(0.6\ \mathrm{dex})^2$, we have removed this branch of the posterior distribution, which is literally unphysical in the sense that none of the data is allowed to be explained by the physical model.

\newpage
\bibliography{paper.bbl}

\end{document}

%% file: table_prior.tex
\begin{table*}[t]\label{tab:prior}\centering
\begin{center}
\begin{tabular}{l | c c} 
\multicolumn{3}{c}{Table 2}\\
\tableline\tableline
Parameter & Definition & Prior Distribution \\
\hline
  $\log (\rho_0 /\cos~i)$ [$M_\odot$ pc$^{-3}$] & Central volume density    & Uniform $(-2.3, 2.3)$ \\
  $\log r_0$ [pc]   & Scale radius              & Uniform $(-2.3, 1)$ \\
  $\log \alpha$ & Envelope steepness        & Uniform $(-2.3, 2.3)$ \\
  $\log R_T$ [pc]   & Truncation radius         & Uniform $(-2.3, 2.3)$ \\
  $\log \sigma^2$ [$(\sunits)^2$] & Unmodeled column density error & Uniform $(-2.3, \log\sigma_{\Sigma_i}^2 - 0.6^2)$\\
\end{tabular}
\end{center}

\end{table*}